# Domain-Wall-Mediated Interfacial Ferroelectric Switching

*Hao-Wen Xu[a], Wen-Cheng Fan[a], Jun-Ding Zheng[a], Cheng-Shi Yao[a], Ni Zhong[a,c,*],*

*Wen-Yi Tong[a,b,*], Chun-Gang Duan[a,c,*]*

[a] Key Laboratory of Polar Materials and Devices (MOE), School of Physics and Electronic Science and Shanghai Center of Brain-inspired Intelligent Materials and Devices, East China Normal University, Shanghai 200241, China

[b] Suzhou Laboratory, 388 Ruoshui Road, Suzhou 215123, People's Republic of China

[c] Collaborative Innovation Center of Extreme Optics, Shanxi University, Taiyuan, Shanxi 030006, China

## ABSTRACT

Interfacial ferroelectricity offers a promising platform for ultrafast, low-power memory devices. While previous studies have demonstrated the importance of domain wall in polarization switching, the coexistence of various domain wall types and their impact on polarization stability lacks fundamental understanding. By integrating first-principles calculations, machine learning methods, and experimental validations, we show that domain walls connect opposite polarization states and respond to out-of-plane electric field through polarization vector deviation, leading to inhomogeneous interlayer sliding and domain-wall migration. This mechanism bears clear resemblance to that in traditional ferroelectrics. Notably, different domain wall types result in distinct switching behaviors, which play a crucial role in determining the reversibility of polarization switching. We then propose strategies beyond ideal conditions to achieve non-volatile ferroelectric switching, which are supported by our experimental observations. These insights shed light on the microscopic switching mechanism in hexagonal interfacial ferroelectrics, offering guidance for future nanoelectronics applications.

## KEYWORDS

Polarization switching mechanisms, Interfacial ferroelectrics, Non-volatile ferroelectricity, Hexagonal moiré superlattice, Machine learning methods, Molecular dynamics

# MAIN TEXT

Ferroelectric materials exhibit reversible spontaneous polarization that can be switched by electric fields, making them promising candidates for memory applications.[1–4] Interfacial ferroelectricity has emerged as a prominent research focus due to its exceptional properties that surpass conventional bulk ferroelectrics.[5–7] In low-dimensional systems, ferroelectricity can be intrinsically induced or stabilized at interfaces through symmetry breaking, charge redistribution, and localized strain fields, enabling robust switchable polarization.[8–13] Notably, such ferroelectricity can persist down to atomic thickness, ideal for miniaturized devices and tunable by strain or external fields.[14–16] Particularly in van der Waals (vdW) structures,[17] interfacial ferroelectricity demonstrates ultralow power consumption,[8,18] ultrafast polarization switching[19] and superior fatigue resistance,[20,21] highlighting strong potential for practical applications.

Among such systems, vdW hexagonal structures have received the most attention due to their ultralow energy barriers and ultrafast switching.[22–26] The full-layer sliding picture from the concept of sliding ferroelectricity[8] faces intrinsic challenges under $C_3$ symmetry, as three equivalent sliding pathways lead to directional ambiguity and ultralow barriers render the sliding termination ill-defined. Recent theoretical[19,27] and experimental[28] studies have shown that $C_3$-symmetry-protected monodomains lack sufficient in-plane (IP) driving force to switch under out-of-plane (OOP) electric field, highlighting the indispensable role of domain walls(DWs)[28,29], where off-diagonal Born effective charges provide the required IP driving force for polarization switching[27,30]. However, nonvolatility, a defining feature of ferroelectricity, has received limited attention in such systems. Thus, how DWs, known to be important in switching dynamics, influence polarization retention remains unexplored.

In this work, combining first-principles calculations with machine-learning methods, we investigate DW-mediated polarization switching in vdW interfacial ferroelectrics and its implications for nonvolatile behavior under an OOP electric field. We systematically analyze the classification of DWs, thereby uncovering the structural compatibility between different DW types. The evolution of their local polarization vectors under electric field—analogous to that in conventional ferroelectrics—is structurally manifested as spatially varying interlayer sliding across the moiré lattice. Importantly, using molecular dynamics simulations of bilayer moiré structures, we investigate how DW characteristics influence ferroelectric retention, revealing the key role of DWs in the polarization volatile behavior. Based on this, we propose practical strategies to realize non-volatile interfacial ferroelectricity, supported by our experimental validation. This work uncovers the critical role of DWs in interfacial ferroelectric switching and non-volatility control, advancing the understanding and application of interfacial ferroelectricity.

Bilayer boron nitride (BN), the simplest hexagonal interfacial ferroelectric, is chosen as a model system to investigate polarization switching in this class of materials. In hexagonal bilayer systems, AB and BA stackings(Figure S1) represent the most energetically stable states, exhibiting opposite OOP polarization and no IP polarization.[8,14,23,31,32] Monodomains composed purely of AB or BA stackings preserve $C_3$ symmetry, ensuring equal atomic displacement probabilities along three IP directions. This symmetry prevents net IP motion and suppresses polarization switching (Movie S1), consistent with our experimental observations in both BN (Figure S17) and $WSe_2$, a typical TMD material (Figure S18), and also aligns with previous reports.[27,28] Therefore, breaking $C_3$ symmetry is crucial for enabling directional IP interlayer motion and achieve polarization reversal.

Unlike monodomain configurations, multidomain systems containing oppositely polarized AB and BA domains provide a feasible switching path through the intermediate SP stacking, which is the metastable state between AB and BA on the energy landscape (Figure S1, S2b).[25,33] SP stacking, as main configuration of DWs, exhibits strong IP polarization but vanishing OOP polarization[34](Figure S2a), arising from the relatively lateral displacement of boron atoms to one side of nitrogen atoms. while its twofold rotational symmetry along the IP direction cancels the OOP component.[19] IP polarization inherently breaks $C_3$ symmetry, making DW responsive to external fields. For both BN and $WSe_2$ (Figures S17 and S18), when an electric field is applied to regions containing DWs, the walls move correspondingly, leading to polarization switching, otherwise no DW motion is observed. These results highlight the essential role of domain walls in vdW interfacial ferroelectric switching, consistent with previous experimental reports.[21,28,32,35] Therefore, we investigate multidomain systems where SP stacking forms DWs bridging reversed polar domains (AB and BA) with equal area, resulting in a moiré domain antiferroelectric (MDAF) arrangement.[35] The application of an OOP electric field drives DW motion, where domains aligned with the field expand while opposite ones contract, generating a net polarization.[36] Reversing the field switches the polarization, enabling global ferroelectric switching.[13,37]

Due to the hexagonal lattice symmetry, DWs can be classified into two characteristic orientations: armchair (AC) and zigzag (ZZ).[38] Each has two variants with IP polarization directions differing by 60°. Due to structural compatibility, two AC-type DWs can coexist within a single lattice period, the same holds for ZZ-type, whereas AC- and ZZ-type walls are mutually exclusive within one unit cell. In AC-type walls, the IP polarization forms 0° and 60° angles with the wall boundary, corresponding to the AC-I and AC-II types respectively; whereas in ZZ-type walls, the angles are 30° and 90° (ZZ-I and ZZ-II) (Figure 1). The anisotropic lattice mismatch is expected to correlate with local polarization direction. In ZZ-II DWs it relaxes perpendicular to the wall, while in AC-I it relaxes parallel to it, giving ZZ-II a larger width. Figure 1 shows a systematic increase in DW width across types: AC-I < ZZ-I < AC-II < ZZ-II, consistent with previous theoretical works[19,39] and unaffected by simulation cell size (Figure S7). The energy differences of various polarization orientations of DW reported previously[39] provides a rational explanation for this trend in DW width.

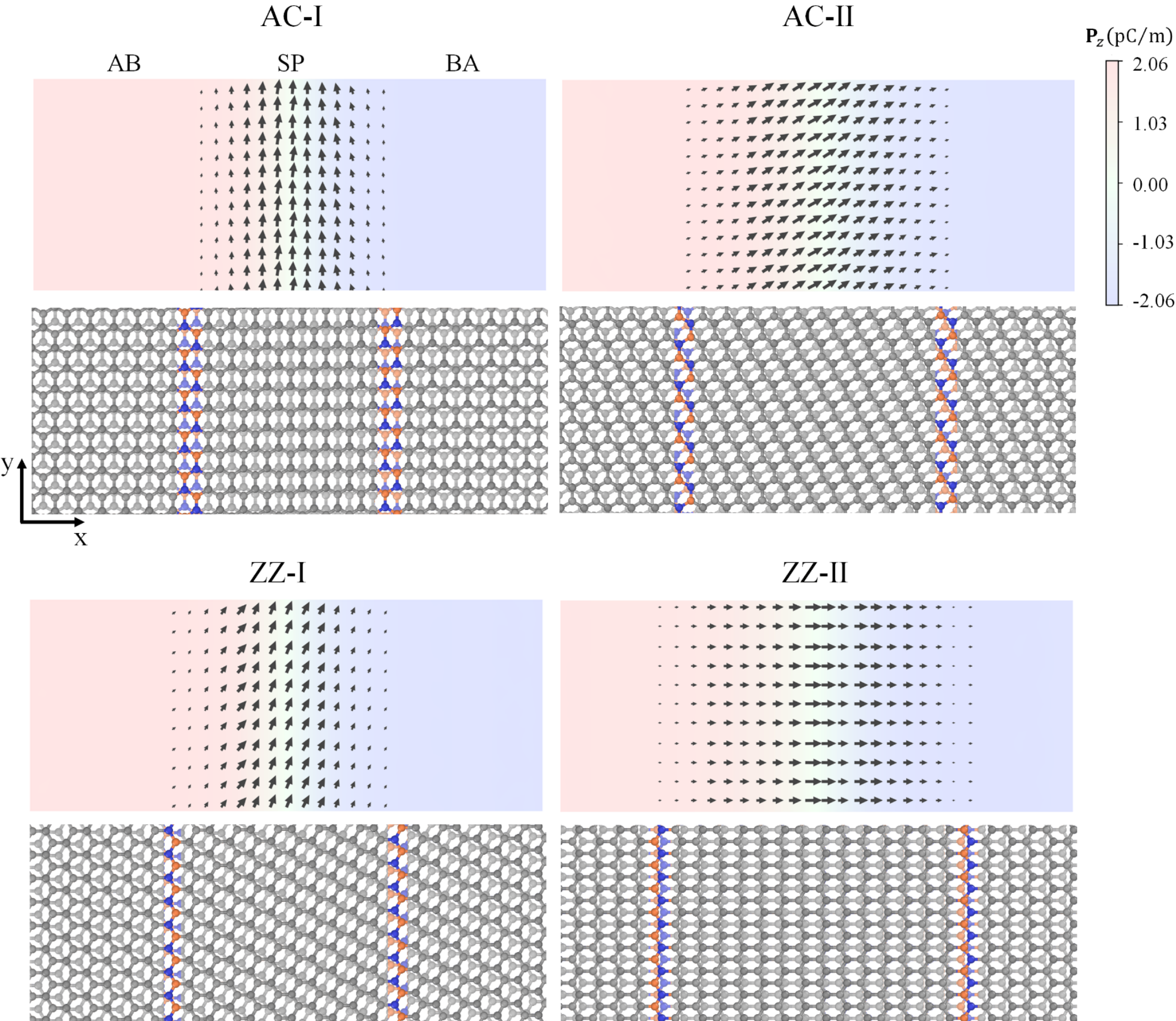

**Figure 1. Top-view atomic structures and corresponding local polarization distributions of four types of DWs.** These DWs are classified according to the direction of IP polarization and the atomic structure within the wall. The top two represent the AC-type, while the bottom two correspond to the ZZ-type. The intensity and hue of red and blue indicate the magnitude and direction of OOP polarization, whereas the size of the dark grey arrows reflects the strength of IP polarization. B atoms are shown as orange spheres, and N atoms as blue spheres.

In conventional ferroelectrics, the emergence of a second polarization component enables the formation of Bloch- and Néel-like DWs.[40–42] Here for multidomain interfacial ferroelectrics, DWs separate AB and BA stacking regions with reversed OOP polarization. Polarization vectors retain nearly constant magnitude across the entire BN structure, including DWs, so the variation arises mainly from directional rotation rather than amplitude change (Figure S2a). Among the four DW types, AC-I DW, similar to $BaTiO_3$,[43] exhibits a Bloch-like polarization profile, where polarization rotates within the wall plane **(**Figure 2a**)**. In contrast, like $Pb(Zr,Ti)O_3$,[44,45] the polarization of ZZ-II DW rotates across the wall along the normal direction, indicating a Néel-like character **(**Figure 2d**)**. Additionally, both AC-II and ZZ-I walls display hybrid rotation patterns combining Bloch- and Néel-like features.

For traditional ferroelectrics with DWs, polarization switching proceeds as follows: upon the application of a negative electric field, the DWs propagate laterally, increasing the area of –P domains while reducing that of the +P domains, known as the (Kolmogorov-Avrami-Ishibashi)KAI model.[46,47] A similar switching mechanism occurs here. When an OOP electric field is applied, local polarization vectors tend to rotate toward the field direction. Vectors that initially lie IP gradually rotate and acquire an OOP component. To achieve this, local structure reconstructs, resulting in reduced IP and enhanced OOP polarization, as clearly illustrated in the dashed boxes near the SP and BA′ regions in Figure 2a. For stacking regions with polarization opposing the OOP field (e.g. AB in Figure 2a), an IP component emerges to weaken the antiparallel

OOP component. It is worth noting that the reorientation of local polarization vectors exhibits a strong correlation with transformations in the stacking configuration: AB → AB', SP → SP', and BA' → BA. All these transformations exhibit the trend of approaching BA configuration, whose polarization aligns with the field. Globally, DWs shift toward AB regions, generating a net polarization paralleled to BA direction. Such behavior, characterized by DW-mediated polarization switching, is analogous to the KAI model in conventional ferroelectrics. Thus, polarization switching here is not necessarily a long-range process requiring global interlayer sliding, but instead arises from the collective effect of local reconstruction. Since these responses mainly involve local stacking rearrangements achieved via interlayer sliding, they can be viewed as local interlayer sliding along the direction of the DW's IP polarization (Figure S12). During this process, local polarization vectors undergo continuous rotation to align with the applied field. Such reorientation preferentially occurs in DWs containing IP polarization components, where local $C_3$ symmetry breaking allows smooth polarization evolution. Furthermore, from the perspective of polarization response to the electric field, the evolution of Bloch-like and Néel-like polarization vectors within DWs still follows the framework of conventional ferroelectric switching.[40,45]

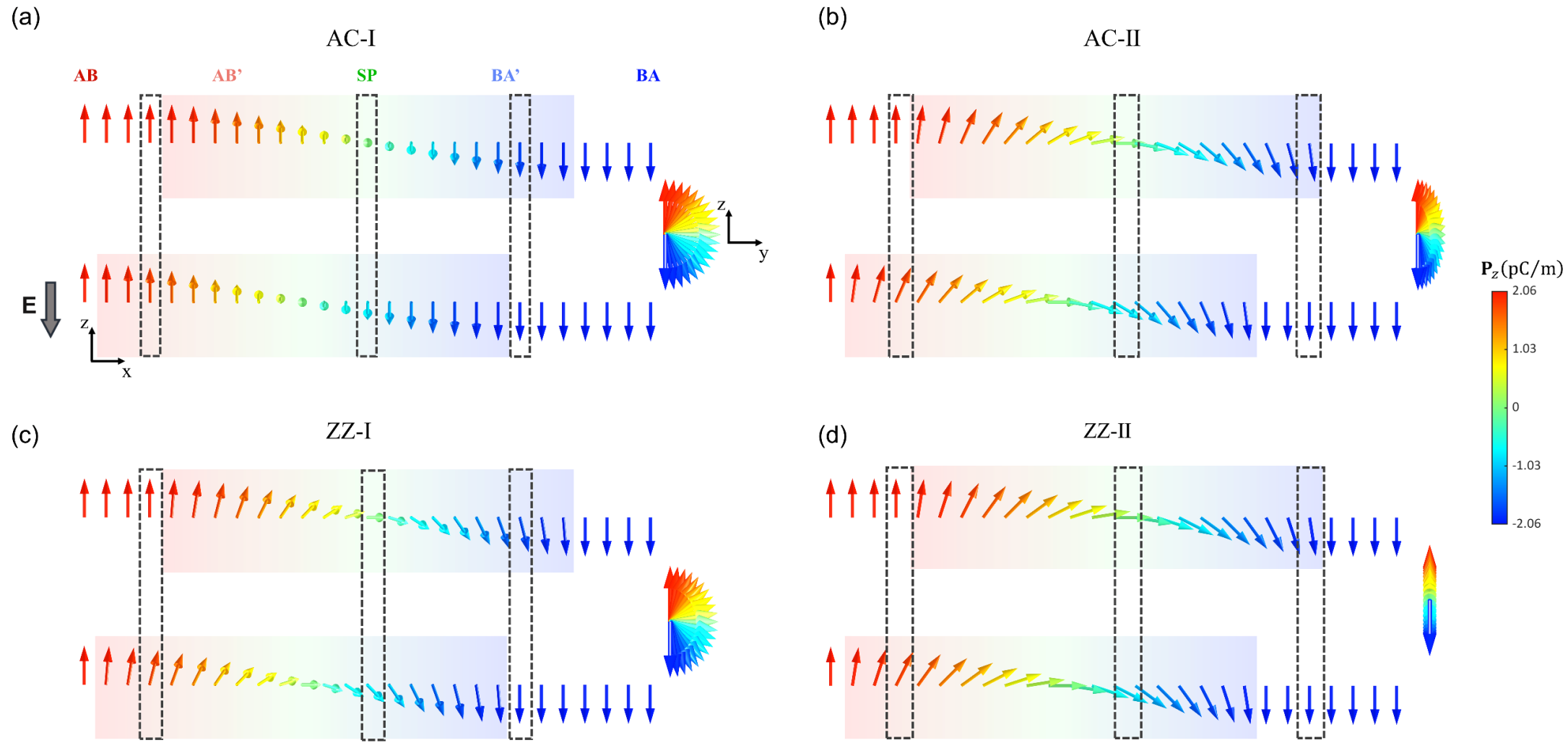

**Figure 2. Changes in local polarization vectors across the four types of DWs(a-d) before and after applying an electric field.** Each panel includes side views along both the *x* and *y* directions. The grey arrow indicates the direction of the applied electric field. The color of the polarization vector arrows represents the magnitude and direction of OOP polarization. The dark grey dashed boxes highlight changes in polarization vectors before and after electric field application within the AB, SP, and BA′ regions, respectively.

In hexagonal interfacial ferroelectrics, multidomain structures can be morphologically categorized into rectangular and triangular types, both intrinsically exhibiting a MDAF state with equally-sized AB and BA polar domains. Rectangular domain structures can be further classified into single-type and mixed-type configurations, depending on the types of DWs present. Structures with single-type DWs contain only one type of DW. When a downward electric field is applied, these DWs migrate and annihilate, transforming the system into a monodomain BA-stacked state with downward polarization(Figure 3a), consistent with previous studies.[19,21] The annihilation processes for all four single-type DW structures are documented in movies S2-S5. In these structures, adjacent DWs share identical polarization directions and structural orientations, enabling them to merge and disappear under the field. Once formed, the monodomain state is in principle irreversible (movie S1), so that removing the electric field cannot restore the multidomain configuration.

As expected, mixed-type DW structures can be classified into AC-type and ZZ-type due to structural compatibility, within both cases the widths of all four DW types remain consistent with those in single-type configurations and are unaffected by periodicity (Figure 3, c-d). Unlike mergeable single-type wall structures, adjacent DWs in mixed-type wall structures possess different polarization directions as well as structural orientations, which limit the extent of DW motion. As shown in Figure 3b, DWs do not merge in mixed-type wall structures, preventing transformation into a monodomain state. Instead, these systems exhibit only limited DW motion. The DWs shift under a downward electric field, reducing the AB domain area while enlarging the BA domain, yielding a net OOP polarization globally (movie S6-S7).

Given the intrinsic energy degeneracy of AB and BA stacking, unequal domain areas induced by the electric field are expected to leave the system's total energy nearly unchanged after field removal, implying potential non-volatile behavior. Surprisingly, our observations show that the system reverts to the MDAF state once the field is withdrawn (Figure 3b and movies S8-S9)**.** Since the relative areas of AB and BA no longer affect total energy, observed domain area changes can be attributed to post-field

DW reconfiguration. DWs primarily consist of SP stacking, which has higher intrinsic energy than both AB and BA. Under an external electric field, the energy difference between SP and BA decreases (Figure S2b), leading to an expansion of SP regions. After removing the field, AB and BA stackings return to energetically degenerate ground states. However, the area of high-energy SP regions increases compared to the pre-field state, prompting DWs to retract and reduce their area to minimize total energy. Calculated DW widths before, during, and after field application confirm this trend. After applying the field, DW widths increase in mixed-type structures (Figure 3c–d), while their relative order remains consistent with single-type configurations throughout the field cycle. Furthermore, to exclude the influence of lattice periodicity, we compared the total DW width for mixed-type DW structures with different unit cell sizes, and in all four periodic cases, DW width increases under the field. Upon field removal, these enlarged SP regions become energetically unfavorable and tend to shrink, driving DW contraction and returning them to original positions.

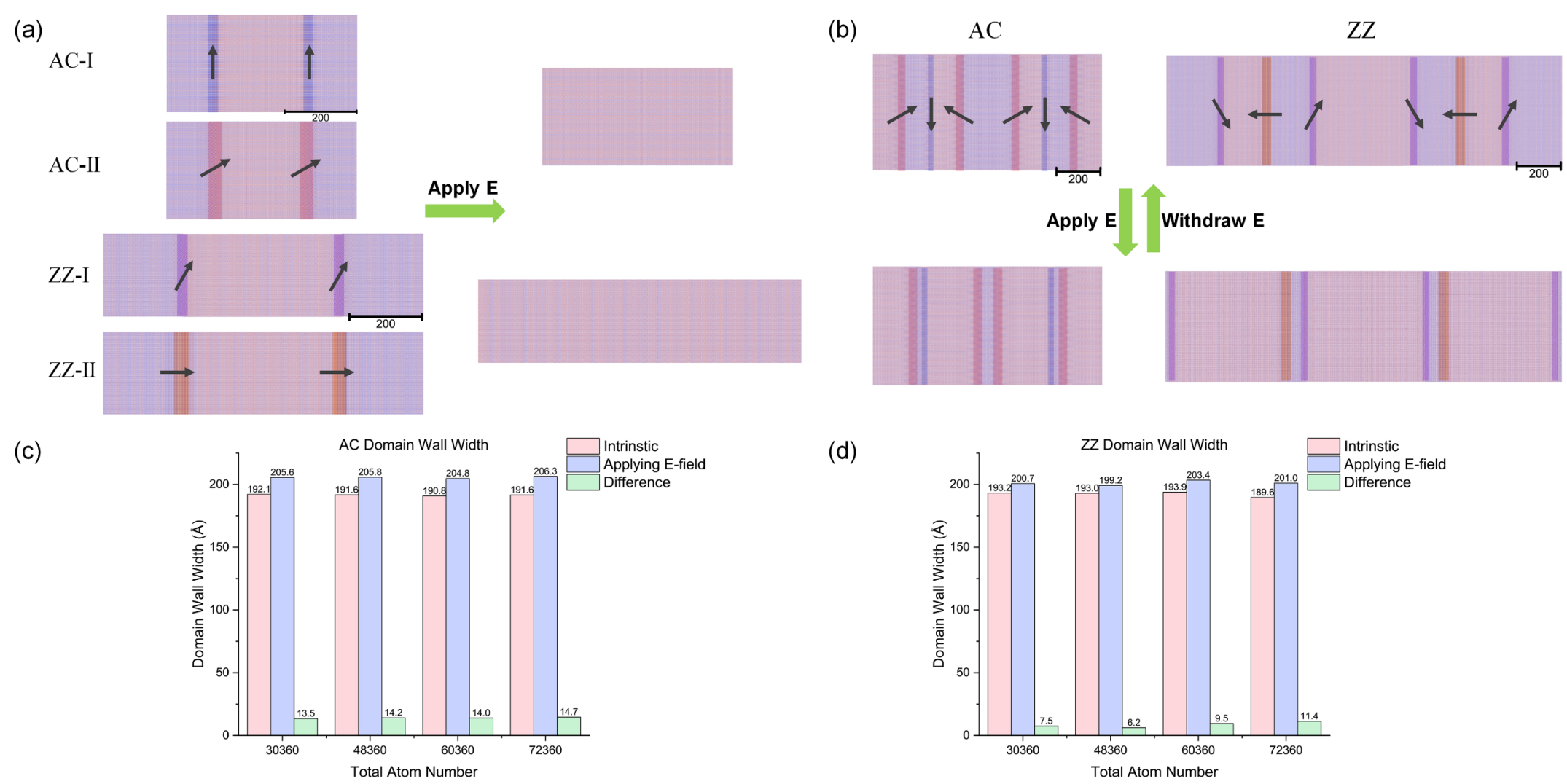

**Figure 3. Changes in domain structure and DW width of the rectangular domain structures before and after electric field application.** (a-b) are overlaid visualizations of the atomic structure and the projected DW area (fig S8). (a) Evolution of single-type DW structures under electric field, from top to bottom: AC-I, AC-II, ZZ-I, and ZZ-II. Pale blue and pink regions indicate local AB and BA stacking, respectively, where B atoms (red spheres) and N atoms (blue spheres) are shown in the atomic representations. The blue, magenta, purple, and orange stripes mark the DW regions of the four types, as identified and quantified following the analysis described in Methods. Dark grey arrow denotes IP polarization directions in the DW. After applying the electric field, all single-type DW structures transform into a single BA-stacked domain. (b) Evolution of mixed-type DW structures before and after applying the electric field. The left panel shows AC DW structure, and the right panel shows ZZ DW structure. After removing the field, the domain structure recovers. Scale bars in (a-b) represent 200 Å. (c-d) Changes in total DW width before and after electric field application for AC-type and ZZ-type mixed DW structures, respectively. To rule out the influence of periodicity on the conclusions, DW widths are investigated across four system sizes, corresponding to primitive cells with 30360, 48360, 60360, and 72360 atoms, respectively.

A similar phenomenon is observed in the response of triangular domains to the electric field. The rotational symmetry of equilateral triangles implies that only one type of DW can exist within each triangular domain. Accordingly, triangular configurations are categorized into four types based on their DW types (Figure 4, a-d). Among them, the AC-I structure is formed purely by interlayer twist, while the ZZ-II type originates from biaxial strain induced by interlayer lattice mismatch, and the AC-II and ZZ-I structures are hybrids involving both twist and strain. Notably, strain-containing structures exhibit distinct rotational distortions in both domains and DWs.

Initially, oppositely polarized AB and BA domains are balanced, forming a MDAF state. Upon electric field application, the system transitions to a ferroelectric state, extensively confirmed by experimental studies[13,24] as well as theoretical analysis.[19,48] As shown in Figure 4a-d and movies S10-S13, all four triangular types (AC-I to ZZ-II) exhibit consistent responses under the field: the DWs become distorted, AB domains shrink, and BA domains expand, resulting in a net OOP polarization aligned with the field. Similar to rectangular domains, triangular domains cannot evolve into a monodomain state owing to the inherent incompatibility among the three DWs, whose intrinsically non-collinear polarization directions render their atomic structures mutually incommensurate. Thus, triangular domain systems can be considered as a special form of mixed-type DW structures.

Apart from width changes, DWs in triangular domains also elongate under the electric field. To quantify these effects, we evaluate the area ratio of DWs, which reflects changes in both width and length. After field removal, DWs revert to their original configuration, AB and BA domains regain equal areas, and the net polarization disappears. Besides, AC-I walls exhibit the smallest area ratio and ZZ-II the largest (Figure 4e), consistent with rectangular domains. Field application increases area ratios across all types due to enhanced distortion. For example, AC-I walls, initially straight, become curved under the field, increasing their length and area despite minimal width change. The other three DW types show similar behavior. Upon removal of the field, these area ratios fall back to their original values. This mechanism explains the volatile ferroelectric response observed in MD simulations of triangle-domain structures (Figure 4a-d and movies S14-S17) and is further validated by our experimental results with AC-I DW (Figure 5a).

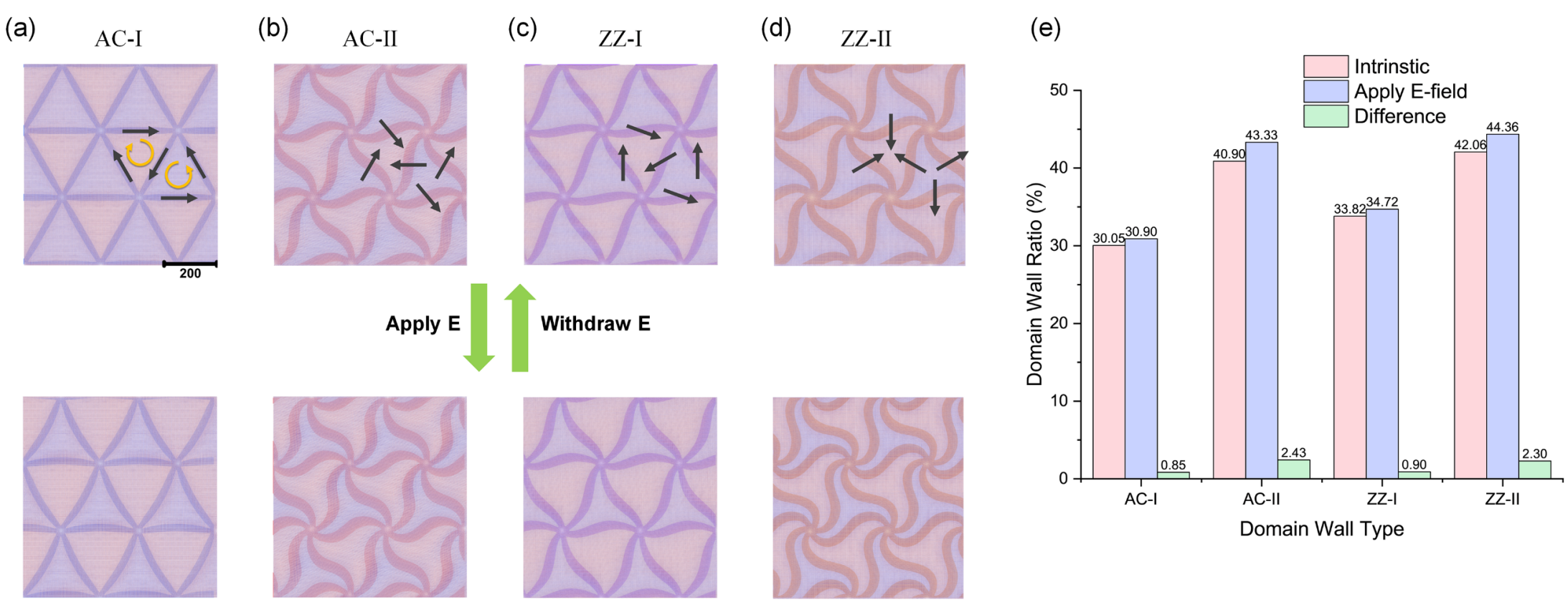

**Figure 4. Changes in domain structure and DW ratio of the triangular domain structures before and after electric field application.** Domain structure evolution of the four DW types — AC-I (a), AC-II (b), ZZ-I (c), and ZZ-II (d) in triangular domain structures— before and after electric field application. Pale blue and pink regions indicate local AB and BA stacking, respectively. The blue, magenta, purple, and orange areas correspond to the DW regions of the four types. As illustrated in Figure S10, (a-d) show superimposed images of atomic structures and DW distributions (Figure S9), representing systems with 53068, 53474, 53132 and 53362 atoms in their respective primitive cells. The selection aimed to maintain comparable system sizes. Scale bars in panels (a-d) each represent 200 Å. (e) Variation in DW ratio for the four DW types before and after electric field application.

Such field-induced reversible behavior closely parallels that observed in rectangular domain structures (Figure 3), suggesting a shared physical origin. The volatility in hexagonal interfacial ferroelectrics stems from the intrinsic energy cost of DWs. Under an electric field, AB domains shrink and BA domains expand, increasing the area fraction of high-energy DWs and raising the system's total energy. To minimize energy, DWs contract after the field is removed, restoring the MDAF state. Although triangular domains contain only one type of DW, the $C_3$ symmetry of their polar structure introduces intrinsic angular mismatch between adjacent walls, preventing their irreversible merging. In contrast, rectangular domains with single-type walls can form monodomain states due to structural compatibility between adjacent DWs.

The above discussion is based on ideal, defect-free, and perfectly periodic structures, whose volatility is likely a common characteristic of vdW interfacial ferroelectrics, as indicated by both theoretical and experimental results in $WSe_2$ (Figures S14 and S16). In real samples, however, edge defects and other deviations from perfect periodicity are inevitable. Non-volatile interfacial ferroelectricity may therefore be achievable in rectangular single-type DW systems, where external fields push DWs to sample boundaries. To shed light on polarization nucleation processes relevant to device applications,[28] we compare the critical nucleus sizes for polarization reversal at the boundaries of the four DW types. The obtained order AC-I $<$ ZZ-I $<$ AC-II $<$ ZZ-II (Figure S11) indicates easier reversal within the formation of AC-I walls. At sample boundaries, defects and local strain fields like bubbles can pin the walls,[21,49] preventing their contraction after field removal and maintaining net OOP polarization. When a reverse electric field is applied, it helps release the DWs from the pinning centers, allowing them to move again and switch the system to the opposite polarization state.

Our experimental observations corroborate this mechanism for achieving non-volatility in interfacial ferroelectrics. In practical sample preparation, non-ideal factors such as mechanical stress, trapped contaminants, and uneven adhesion can introduce wrinkles and interfacial bubbles, disrupting the moiré periodicity and leading to irregular domains (Figure 5b). Piezoresponse force microscopy (PFM) amplitude images show that DWs initially located near the sample boundaries (indicated by black dashed lines). Under an electric field, AB domains contract while BA domains expand until DWs migrated to the opposite boundary, completing polarization reversal. Eventually, DWs became pinned at that boundary, stabilizing the reversed polarization and enabling non-volatile polarization control, suggesting the feasibility of our theoretical strategy.

The previous discussion indicates that volatile behavior in interfacial ferroelectrics can be attributed to the expansion of the DW area under the influence of electric field. Based on this understanding, to achieve non-volatility, such expansion should be suppressed. Inspired by the strain-engineering approach[50] that distorts equilateral domains into elongated ones to achieve full local control, resembling a zipper-like modulation of the domain network, here we extend this method in pursuit of non-volatility. Through this targeted strain modulation, part of the DWs can be merged into a single wall. This reduction in DW area lowers the system's energy after field removal, preventing reversion to the MDAF state. Upon application of a reverse field, the merged walls separate while others combine sequentially, completing switching in a "zipping" and "unzipping" process. This alternating wall closure and separation suggest a possible route toward non-volatile polarization control, as further supported by our experimental work under review.

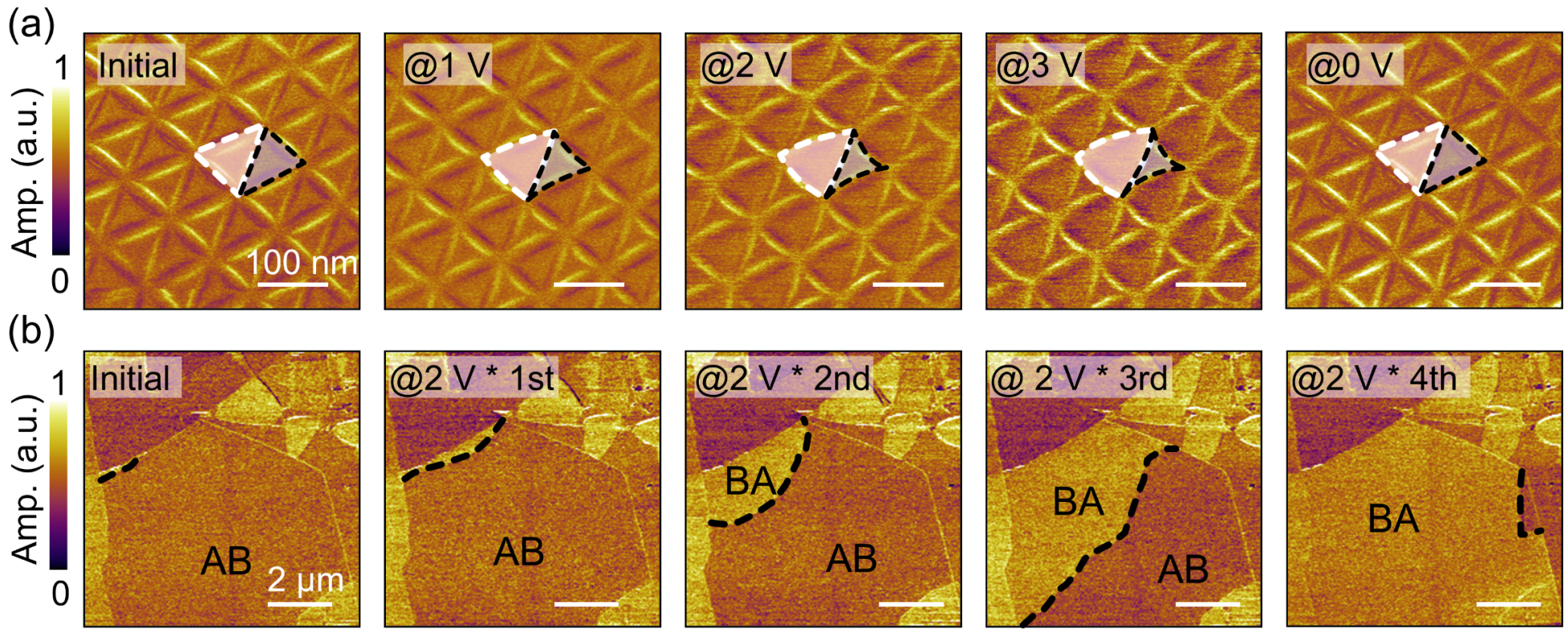

**Figure 5. Dynamic demonstration of DW movement under electric field.** (a) PFM amplitude image of DW movement in triangular moiré domains under electric field. (b) PFM amplitude image of DW movement in irregular moiré domains under electric field.

Our work investigates the important role of domain walls in polarization switching and its nonvolatility in hexagonal interfacial ferroelectrics. Under an OOP electric field, analogous to conventional ferroelectrics, local polarization vectors near domain walls reorient toward the field direction, inducing local stacking reconfigurations via inhomogeneous interlayer sliding. Macroscopically, this manifests as domain wall motion and the growth of field-aligned domains, indicating that polarization reversal can proceed through localized polarization responses around domain walls rather than long-range full-layer sliding. Through MD simulations, we track how different types of moiré structures evolve during the application and withdrawal of electric field. Specifically, the expansion of domain wall area is identified as a critical factor of volatility in mixed-type systems, posing a challenge to robust ferroelectric retention. However, by exploiting beyond-ideal systems containing either single or mixed-type walls, feasible routes toward non-volatile polarization control are explored, further supported by our experimental results. Our findings reveal the pivotal role of domain wall in interfacial ferroelectric switching and non-volatile control, deepening the understanding of interfacial ferroelectricity and offering valuable insights for the design of next-generation electronic devices.

## ASSOCIATED CONTENT

### Supporting Information

Methods contains:

- First-principles calculations
- Machine learning
- DW width calculations
- Sample fabrication
- PFM measurement

19 figures (S1-S19) showing:

- Various stacking configurations and their polarization/energy characteristics
- Evolution of energy/force and polarization models
- Atomic structures of AB/SP/BA stackings
- DW characterization and behavior under electric fields
- Analysis of rectangular and triangular domain structures
- Experimental evidence highlighting the importance of DWs in polarization switching
- Microscopic visualization of local sliding processes
- Theoretical and experimental validation based on a representative TMD material, $WSe_2$

Two large MP4 files, containing the 17 supplementary movies (S1–S17) for ease of viewing, demonstrate:

- Molecular dynamics simulations of monodomain structures
- DW dynamics under applied electric fields
- Recovery processes after field removal
- Both rectangular and triangular domain systems

## AUTHOR INFORMATION

### Corresponding Authors

**Ni Zhong** - Key Laboratory of Polar Materials and Devices (MOE), School of Physics and Electronic Science and Shanghai Center of Brain-inspired Intelligent Materials and Devices, East China Normal University, Shanghai 200241, China; Collaborative Innovation Center of Extreme Optics, Shanxi University, Taiyuan, Shanxi 030006, China; orcid.org/0000-0002-6875-4007; Email: nzhong@ee.ecnu.edu.cn

**Wen-Yi Tong** - Key Laboratory of Polar Materials and Devices (MOE), School of Physics and Electronic Science and Shanghai Center of Brain-inspired Intelligent Materials and Devices, East China Normal University, Shanghai 200241, China; Suzhou Laboratory, 388 Ruoshui Road, Suzhou 215123, People's Republic of China; orcid.org/0000-0003-4910-951X; Email: wytong@ee.ecnu.edu.cn

**Chun-Gang Duan -** Key Laboratory of Polar Materials and Devices (MOE), School of Physics and Electronic Science and Shanghai Center of Brain-inspired Intelligent Materials and Devices, East China Normal University, Shanghai 200241, China; Collaborative Innovation Center of Extreme Optics, Shanxi University, Taiyuan, Shanxi 030006, China; orcid.org/0000-0002-5380-4980; Email: cgduan@clpm.ecnu.edu.cn

**Author Contributions**

W.Y.T and C.G.D conceived the idea and supervised the work. H.W.X carried out first-principles calculations, machine learning MD simulations, and did the data analysis. W.C.F performed the PFM experiments supervised by N.Z. H.W.X wrote the paper with input from other co-authors. All the authors reviewed and modified the manuscript. H.W.X and W.C.F contributed equally to this work.

**Funding Sources**

This work was supported by the National Key Research and Development Program of China (Grants No. 2022YFA1402902 and No. 2021YFA1200700), the National Natural

Science Foundation of China (Grants No. 12134003 and No. 12304218), Shanghai Science and Technology Innovation Action Plan (Grant No. 21JC1402000), Shanghai Pujiang Program (Grant No. 23PJ1402200), and East China Normal University Multifunctional Platform for Innovation.

**Notes**

Authors declare that they have no competing interests.